\documentclass[twocolumn,preprintnumbers, superscriptaddress, reprint, longbibliography, prl]{revtex4-1}
\usepackage{amsmath}
\usepackage{stmaryrd}
\usepackage{txfonts}
\usepackage{amssymb}
\usepackage{mathrsfs}
\usepackage{graphicx}
\usepackage{dcolumn}
\usepackage{bm}
\usepackage{braket}
\usepackage{epsfig}
\usepackage{color}
\usepackage{ulem}
\usepackage[resetlabels,labeled]{multibib}

\newcommand{\ys}[1] {{\color{black} #1}}

\usepackage[colorlinks=true, linkcolor=blue, citecolor=blue, urlcolor=blue]{hyperref}
\begin{document}
\preprint{\href{https://doi.org/10.1103/PhysRevLett.125.226401}{Y. Su and S.-Z. Lin, Phys. Rev. Lett. {\bf 125}, 226401 (2020).}}


\title{\ys{Current-induced reversal of anomalous Hall conductance in twisted bilayer graphene}}
\author{Ying Su}
\affiliation{Theoretical Division, T-4 and CNLS, Los Alamos National Laboratory, Los Alamos, New Mexico 87545, USA}
\author{Shi-Zeng Lin}
\affiliation{Theoretical Division, T-4 and CNLS, Los Alamos National Laboratory, Los Alamos, New Mexico 87545, USA}

\begin{abstract}
It is observed experimentally that the sign of the Hall resistance can be flipped by a dc electric current in the twisted bilayer graphene (TBG) at 3/4 filling of the four-fold degenerate conduction flat  bands. The experiment implies a switching of the valley polarization (VP) and topology in TBG. Here we present a theory on the current-induced switching of VP and topology. The presence of current in the bulk causes {the} redistribution of {electron occupation} in bands near the Fermi energy, which then deforms and shifts the band dispersion due to the Coulomb interaction. Above a critical current, the original occupied and empty bands can be swapped resulting in the switching of VP and topology.
\end{abstract}

\date{\today}
\maketitle

{\emph{Introduction.}---The electronic band structure is modified significantly compared to that in a single layer graphene, when two layers of graphene are stacked together \cite{RevModPhys.81.109}. The bandwidth can be controlled by the misalignment angle between two layers \cite{PhysRevLett.99.256802, bistritzer_moire_2011}.} At certain twisted angles called magic angles, the energy bands near the Fermi surface 
{are} extremely flat \cite{bistritzer_moire_2011}, where the Coulomb interaction becomes dominant over the kinetic energy of electrons. It is expected that novel quantum states enabled by the strong electronic correlation will emerge. Indeed, the correlated insulating state and superconductivity have been observed 
{experimentally in the magic-angle twisted bilayer graphene (TBG)} \cite{cao_correlated_2018,cao_unconventional_2018}. Furthermore, the insulating state can have nontrivial topology which is manifested by the intrinsic quantum anomalous Hall effect (QAHE) \cite{Sharpe605,youngscience}. The twisted multilayer graphene therefore becomes an important platform to explore the physics of strong electronic correlation and topology 
\cite{cao_correlated_2018,cao_unconventional_2018,lu_superconductors_2019,Sharpe605,youngscience, Kim3364,PhysRevLett.121.037702,PhysRevLett.121.087001,PhysRevX.8.031089,Fuprx,PhysRevX.8.031088,PhysRevLett.121.217001,PhysRevB.98.075154,PhysRevX.8.041041,PhysRevB.98.235158,guinea2018electrostatic,xie2018nature,Yankowitz1059,kerelsky2019maximized,cao2019strange,polshyn2019large,
chen2019tunable,Codecidoeaaw9770,xie2019spectroscopic,jiang2019charge, choi2019electronic, lu2019superconductors,PhysRevLett.123.046601,shen2019observation,liu2019spin,cao2019electric,
you2019superconductivity,PhysRevB.99.094521,PhysRevLett.122.026801,PhysRevB.98.195101,PhysRevLett.121.146801,PhysRevLett.122.106405,PhysRevX.9.021013,PhysRevLett.123.036401,PhysRevB.99.035111,PhysRevB.98.205151,PhysRevLett.122.246401,PhysRevLett.122.246402,PhysRevB.100.085136,PhysRevB.98.220504,he_giant_2019,PhysRevLett.122.257002,PhysRevB.98.241412,PhysRevLett.121.257001,chatterjee2019symmetry,PhysRevB.99.165112,PhysRevB.99.195114,PhysRevB.99.220507,wu2019ferromagnetism,PhysRevB.99.075127,PhysRevLett.122.016401,lee2019theory,PhysRevB.100.024421,PhysRevX.9.031049,xie2019topology,julku2019superfluid,PhysRevLett.123.237002,liu2019anomalous,saito2019decoupling,ren2019spectroscopic,li2019experimental,PhysRevResearch.1.033126,bultinck2019anomalous,PhysRevX.9.031021,wu2019collective,PhysRevB.101.041112,PhysRevB.101.041113,PhysRevB.101.041410,zhu2020curious}. 

The QAHE with orbital ferromagnetism was observed at 3/4 filling of the {upper flat} bands in TBG recently \cite{Sharpe605, youngscience}. 
{In the absence of electronic interaction,} the {electronic} bands {of TBG} have the four-fold degeneracy associated with the valley and spin degrees of freedom. {The strong electronic interaction in flat bands lifts the valley and spin degeneracy, that results in the valley-spin-polarized insulating state responsible for the QAHE at 3/4 filling \cite{PhysRevB.99.075127,xie2018nature,bultinck2019anomalous,PhysRevResearch.1.033126,PhysRevX.9.031021,wu2019collective}.}
The {nontrivial topology of the} insulating state is characterized by the  Chern number $C=\pm 1$, {where the sign of Chern number depends on which valley is polarized}.  {Remarkably, as demonstrated in experiments, the sign of Hall conductance can be flipped by a dc current, that indicates the switching of Chern number and valley polarization (VP) \cite{Sharpe605,youngscience}.} \ys{{Several theoretical pictures are proposed to explain the experiments} \cite{youngscience, he_giant_2019,huang_current_2020}.} 

In this paper, we propose a mechanism of switching of VP and Chern number in TBG by electric currents in the bulk. The current causes the redistribution of the electron occupation in otherwise fully occupied or empty bands, which deforms and shifts the band dispersion with respect to the Fermi energy $E_F$ due to the Coulomb interaction. Above a threshold current, the major part of the originally empty (occupied) band when current is absent is pushed below (above) $E_F$. The bands then are swapped after relaxation when the current is removed, that hence results in flipping of the VP and topology.

\emph{Toy model.}---{To facilitate the understanding of switching of VP by electric currents,} we use a simple toy model to demonstrate the physical picture. Here we consider a two-band model described by the Hamiltonian
\begin{align}\label{eq1}
\mathcal{H}=\sum_{k,\tau}(\varepsilon_{k,\tau}-\lambda\partial_k \varepsilon_{k,\tau})c_{k,\tau}^\dagger c_{k,\tau}+V\sum_{k_1,k_2}c_{k_1,+}^\dagger c_{k_1,+} c_{k_2,-}^\dagger c_{k_2,-}, 
\end{align}
where $\tau=\pm$ 
{is the effective valley index labeling the two energy bands and} $V$ is the repulsive intervalley interaction. 
The model has time-reversal symmetry (TRS), $\varepsilon_{k,\tau}=\varepsilon_{-k,-\tau}$. The current in the system can be introduced by the Lagrangian multiplier $\lambda$, which \ys{can be derived by considering a system with a threading magnetic flux \cite{supplement}. It can also be understood intuitively in terms of the semi-classical Boltzmann transport theory \cite{supplement}}. The {self-consistent} mean-field Hamiltonian is $\mathcal{H}_{MF}=\sum_{k,\tau} E_{k,\tau} c_{k,\tau}^\dagger c_{k,\tau}$ with $E_{k,\tau}=\varepsilon_{k,\tau}-\lambda\partial_k \varepsilon_{k,\tau}+\Delta_{-\tau}$ and $\Delta_{\tau}=V\sum_k f(E_{k,\tau})$, where $f(E_{k,\tau})$ is the Fermi distribution function.

{Consider the schematic band dispersions shown in Fig. \ref{fig1}(a). The band with valley $+$ is above that with valley $-$ as a consequence of the spontaneous symmetry breaking due to the intervalley interaction.  The system is an insulator at half filling. To introduce current, a fraction of electrons in the lower band must be pumped into the upper band. Because the electron self-energy depends on the electron occupation, the part of $-$ band with group velocity opposite to the current direction rises while the $+$ band shifts downwards. The majority of the $+$ band can be below the $-$ band above a {critical} current, as shown in Fig. \ref{fig1}(b). {After removing the current, the system then relaxes into a state with $+$ ($-$) band occupied (empty), and thus the VP is switched, {as shown in Fig. \ref{fig1}(c)}.

\begin{figure}[t]
  \begin{center}
  \includegraphics[width=8.5 cm]{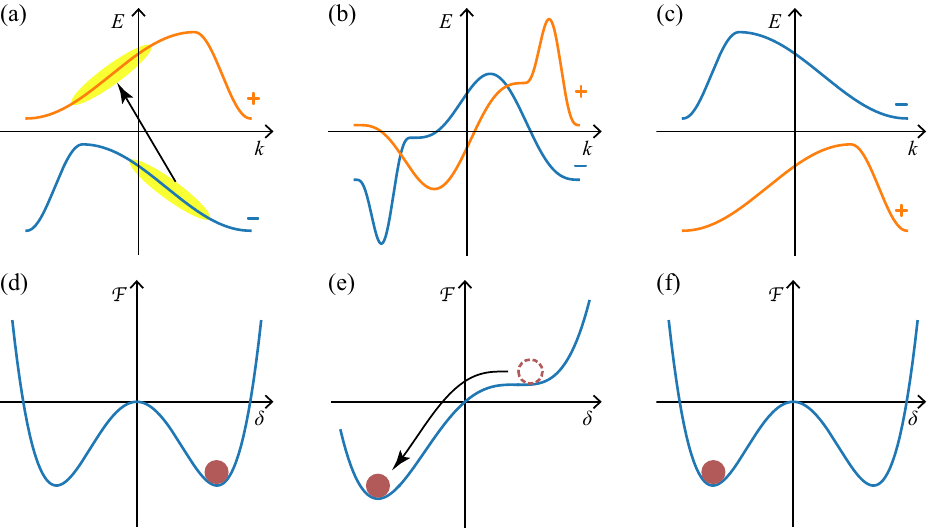}
  \end{center}
\caption{(a)-(c) Schematic band structures for $\lambda=0$, $\lambda\geq \lambda_c$, and $\lambda=0$, respectively. (d)-(f) Free energy profiles correspond to (a)-(c). The two energy bands are labeled by the effective valley index $\pm$. (a) and (c) represent the two degenerate ground states with opposite VP.  (b) is obtained from (a) by introducing the current, and (c) is obtained from (b) by removing the current. The current induced pumping of electrons from the lower band to the upper band is indicated by the arrow and is highlighted in (a). The switching of VP and vanishing of energy barrier in the presence of current is exhibited in (e).      
} 
  \label{fig1}
\end{figure}

{To describe the phase transition associated with the switching of VP,} we introduce {the} VP order parameter $\delta=(\Delta_+-\Delta_-)/2$. Close to the transition temperature $T_c$ when $\delta$ is small, the Ginzburg-Landau free energy of the system can be {expanded} as \cite{supplement}
\begin{align}\label{eq2}
\mathcal{F}=\mathcal{F}_0+\alpha_1\delta+\alpha_2\delta^2+\alpha_3\delta^3+\alpha_4\delta^4.
\end{align}
{Because} both current and $\delta$ are odd under time reversal, it {requires} $\alpha_{1,3}(\lambda)=-\alpha_{1,3}(-\lambda)$ and $\alpha_{2,3}(\lambda)=\alpha_{2,4}(-\lambda)$.  Under the inversion operation {\it{within}} each valley $\hat{\Gamma}_v$, i.e. $\varepsilon_{k,\tau}\rightarrow \varepsilon_{-k,\tau}$, $\delta$ is even while the current is odd. If the system has the symmetry associated with $\hat{\Gamma}_v$, then $\alpha_{1,3}(\lambda)=\alpha_{1,3}(-\lambda)$, which immediately implies $\alpha_{1,3}(\lambda)=0$. Therefore this intravalley inversion symmetry must be broken, i.e. $\varepsilon_{k,\tau}\neq \varepsilon_{-k,\tau}$, in order to switch the valley polarization by current.

It is clear from {Eq. (\ref{eq2})} that the system has two degenerate {ground} states with opposite VP{, which are separated by an energy barrier, as shown in Figs. \ref{fig1}(d) and \ref{fig1}(f)}. The presence of current lifts the degeneracy by increasing the energy of one {valley-polarized state} and reducing the energy of the {other one, as shown in Fig. \ref{fig1}(e)}. Above a {critical} current, the energy barrier vanishes. {The VP of the system is switched if it is initiated at a state, that becomes unstable under the current [see Fig. \ref{fig1}(e)].} On the other hand, if the system sits in {the other state} that remains the energy minimal under the current, {there is} no switching of VP. Thus the switching of VP depends on the direction of current.  }



\begin{figure}[t]
  \begin{center}
  \includegraphics[width=8.5 cm]{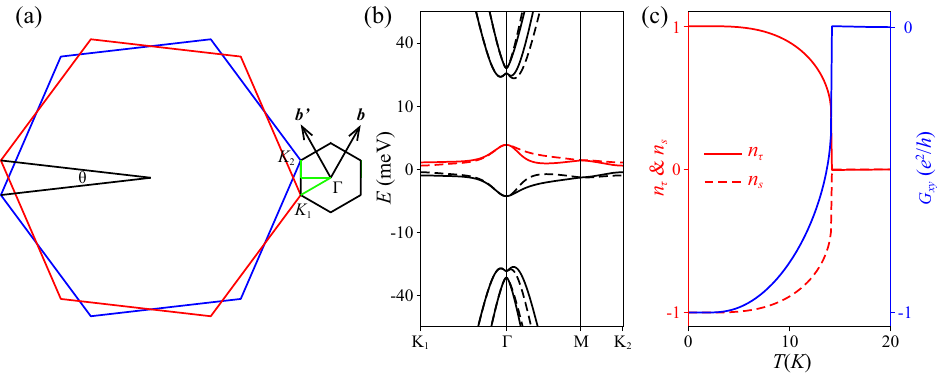}
  \end{center}
\caption{(a) Schematic BZs of the TBG (black hexagon) and of two graphene layers (red and blue hexagons). (b) Low-energy band structure of the TBG along the high symmetry path. The solid and dashed energy bands are from the $\pm$ K valleys, respectively. The upper flat bands are marked by red color.  (c) $n_\tau$ (red solid line), $n_s$ (red dashed line), and Hall conductance $G_{xy}$ (blue solid line) are plotted as a function of the temperature $T$.
} 
  \label{fig2}
\end{figure}

\emph{TBG.}---We then apply our physical picture to TBG with a small twist angle. In TBG, no symmetry guarantees {$ \varepsilon_{\bm{k},\tau}= \varepsilon_{-\bm{k},\tau}$} and therefore it is possible to switch the VP by electric currents. We employ the continuum model Hamiltonian \cite{bistritzer_moire_2011} 
\begin{equation}\label{HTBG}
\mathcal{H}_{\bm{k},\tau} =
\begin{pmatrix}
h_{1,\tau}(\bm{k}) & T_\tau(\bm{r}) \\
T_\tau^\dagger(\bm{r}) &  h_{2,\tau}(\bm{k})   
\end{pmatrix},  
\end{equation}
where $\tau=\pm$ is the valley index and $h_{l,\tau}(\bm{k})$ is the low-energy effective Dirac Hamiltonian of the $l$-th layer graphene.  Under the twist that the first (second) layer is rotated by $\theta/2$ ($-\theta/2$), the Dirac Hamiltonian becomes 
\begin{equation}
h_{l,\tau}(\bm{k}) =\mathcal{R}(\pm\tau\theta/2) \hbar v_F(\bm{k}-\tau\bm{K}_l)\cdot \bm{\sigma}_{\tau} \mathcal{R}(\pm\tau\theta/2)^{-1},
\end{equation}
where $\mathcal{R}(\theta)=e^{-i\theta\sigma_z/2}$ is the rotation operator  and $\bm{\sigma}_\tau=(\tau\sigma_x, \sigma_y)$ is the Pauli matrix for sublattice degree of freedom. 
$\bm{K}_l$ are the corners of Moir{\'e} Brouillon zone (BZ) as shown in Fig. \ref{fig2}(a). 
The interlayer coupling is described by 
\begin{equation}
\begin{split}\label{eq12}
&T_\tau(\bm{r})=T_\tau^{(0)} + e^{-i\tau\bm{b}\cdot\bm{r}} T_\tau^{(1)} + e^{-i\tau\bm{b}'\cdot\bm{r}} T_\tau^{(2)}, \\
&T_\tau^{(n)}=w_{AA}\sigma_0+w_{AB}\cos\left(\frac{2\pi n}{3}\right)\sigma_x+\tau w_{AB}\sin\left(\frac{2\pi n}{3}\right)\sigma_y,
\end{split}
\end{equation}
where $\bm{b}$ ($\bm{b}'$)  are the primitive reciprocal lattice vectors of the TBG   [see Fig. \ref{fig2}(a)]. Due to the lattice relaxation, the interlayer tunneling strength is different for the AA and AB stacking regions, 
{where} $w_{AA}=79.7$ meV and $w_{AB}=97.5$ meV \cite{Fuprx}. The Fermi velocity is $\hbar v_F/a=2.1354$ eV where $a=0.246$ nm is the lattice constant of single layer graphene. $\mathcal{H}_{\bm{k},\pm}$ of the $\pm K$ valleys are related by TRS.
In the experiments, the TBG with $\theta=1.15^\circ$ is  aligned with a hBN substrate \cite{Sharpe605,youngscience} that induces a sublattice potential onto the first layer graphene as $h_{1,\tau}\rightarrow h_{1,\tau} + \Delta_{AB}\sigma_z/2$ \ys{through the interlayer coupling}. We choose the sublattice potential $\Delta_{AB}=20$ meV in this study.  \ys{The Hamiltonian Eq. (\ref{HTBG}) is expressed in the real space representation with $\bm{k}=-i\nabla_{\bm{r}}$ being the momentum operator. Under the plane wave expansion and truncation of reciprocal space to the fourth shell,} 
the low-energy bands of the TBG are \ys{well converged and} shown in Fig. \ref{fig2}(b), where the energy bands from the $\pm K$ valley are represented by solid and dashed lines, respectively. There are eight flat bands including the degenerate spin degree of freedom around charge neutrality. The lattice relaxation isolates the flat bands from remote bands and the \ys{substrate-induced} sublattice potential gaps the upper flat bands from the lower flat bands \ys{by breaking the $\mathcal{C}_2\mathcal{T}$ symmetry that enables the nonzero valley Chern number \cite{PhysRevX.8.031089}. Here $\mathcal{C}_2$ and $\mathcal{T}$ denote the two-fold rotation and time-reversal symmetries, respectively.}

{Here we focus on the 3/4 filling of  the upper flat bands  [the red bands in Fig. \ref{fig2}(b)], where the QAHE emerges \cite{Sharpe605,youngscience}.} 
{By projecting the Coulomb interaction onto the upper flat bands, the Hamiltonian of TBG becomes}
\begin{equation}\label{H0}
\mathcal{H}_0 = \sum_{\bm{k},\tau,s} \left( \varepsilon_{\bm{k},\tau} - \mu \right) c_{\bm{k},\tau,s}^\dagger c_{\bm{k},\tau,s}  + \frac{1}{2A}\sum_{\bm{q}} \rho(\bm{q})V(\bm{q})\rho(-\bm{q}),
\end{equation}
where $\varepsilon_{\bm{k},\tau}$ denotes the spin-degenerate bare dispersion for the  upper flat bands given by $\mathcal{H}_{\bm{k},\tau}\ket{\psi_{\bm{k},\tau}}=\varepsilon_{\bm{k},\tau}\ket{\psi_{\bm{k},\tau}}$, $\mu$ is the chemical potential, and $A$ is the area of the system. 
$s=\uparrow$ or $\downarrow$ represents the spin degree of freedom. $\rho(\bm{q}) = \sum_{\bm{k},\bm{k}',\tau,s}\braket{\psi_{\bm{k},\tau}|e^{i\bm{q}\cdot\bm{r}}|\psi_{\bm{k}',\tau}} c_{\bm{k},\tau,s}^\dagger c_{\bm{k}',\tau,s}$ is the density operator, and
$V(\bm{q})=e^2 \tanh(|\bm{q}|d)/2\epsilon |\bm{q}|$ is the screened Coulomb potential, where  $\epsilon$ is the dielectric constant and $d$ is the distance between TBG and metallic gates. Here we take $d=40$ nm following the experiment \cite{youngscience}.

With the self-consistent Hartree-Fock approximation, the dispersion of the upper flat bands are corrected by the Coulomb interaction as 
\begin{equation}\label{sf}
\begin{split}
&E_{\bm{k},\tau,s} = \varepsilon_{\bm{k},\tau} - \mu\\
&  + \frac{1}{A}\sum_{\bm{q},\bm{k}',\tau',s'} \braket{\psi_{\bm{k},\tau}|e^{i\bm{q}\cdot\bm{r}}|\psi_{\bm{k},\tau}}V(\bm{q}) \braket{\psi_{\bm{k}',\tau'}|e^{-i\bm{q}\cdot\bm{r}}|\psi_{\bm{k}',\tau'}} f(E_{\bm{k}',\tau',s'})\\
& - \frac{1}{A}\sum_{\bm{q},\bm{k}'} \braket{\psi_{\bm{k},\tau}|e^{i\bm{q}\cdot\bm{r}}|\psi_{\bm{k}',\tau}}V(\bm{q})\braket{\psi_{\bm{k}',\tau}|e^{-i\bm{q}\cdot\bm{r}}|\psi_{\bm{k},\tau}} f(E_{\bm{k}',\tau,s})
\end{split}.
\end{equation}
By solving the self-consistent equation, we get four degenerate valley-spin-polarized ground states at 3/4 filling. In the ground states,  one of the four flat bands is above the other three that results in an 
insulating state \ys{when $\mu$ is tuned to be inside the gap}. The experimentally measured energy gap is $\Delta/k_B\approx 27K$ \cite{youngscience},  that can be reproduced by our approach with $\epsilon=58.6\epsilon_0$, where $\epsilon_0$ is the vacuum permittivity.  \ys{The energy gap depends on the  gate distance $d$. As $d$ decreases, the screening of the Coulomb interaction is enhanced and the energy gap is reduced \cite{supplement}.}  

{We define} $n_\tau=\sum_s (n_{+,s}-n_{-,s})$ and $n_s=\sum_\tau (n_{\tau,\uparrow}-n_{\tau,\downarrow})$ to characterize the valley and spin polarization, respectively. $n_{\tau,s}$ is the occupation number of the flat band with the valley-spin indices $(\tau,s)$. 
{We consider a symmetry-breaking state with $n_\tau=-n_s=1$ at zero temperature, as shown in Fig. \ref{fig3}. The four flat bands have nonzero Chern numbers $C=\pm 1$, which are opposite for $\pm K$ valleys as a consequence of TRS. Therefore, the {VP} at 3/4 filling ensures the QAHE, which is manifested by the quantized Hall conductance $G_{xy}=\pm e^2/h$ as observed in experiments \cite{youngscience}}. Because the Hamiltonian Eq. (\ref{H0}) has the spin SU(2) symmetry and valley U(1) symmetry, the long-range spin order is destroyed by thermal fluctuations according to the Mermin-Wagner theorem, while the long-range valley order is allowed \cite{PhysRevLett.17.1133}. 
As temperature increases, the mean field $n_\tau$, $n_s$, and $G_{xy}$ vanish together above a critical temperature $T_c=14.1K$, as shown in Fig. \ref{fig2}(c). \ys{Above $T_c$, the system becomes a simple metal.}


\begin{figure}[t]
  \begin{center}
  \includegraphics[width=8.5 cm]{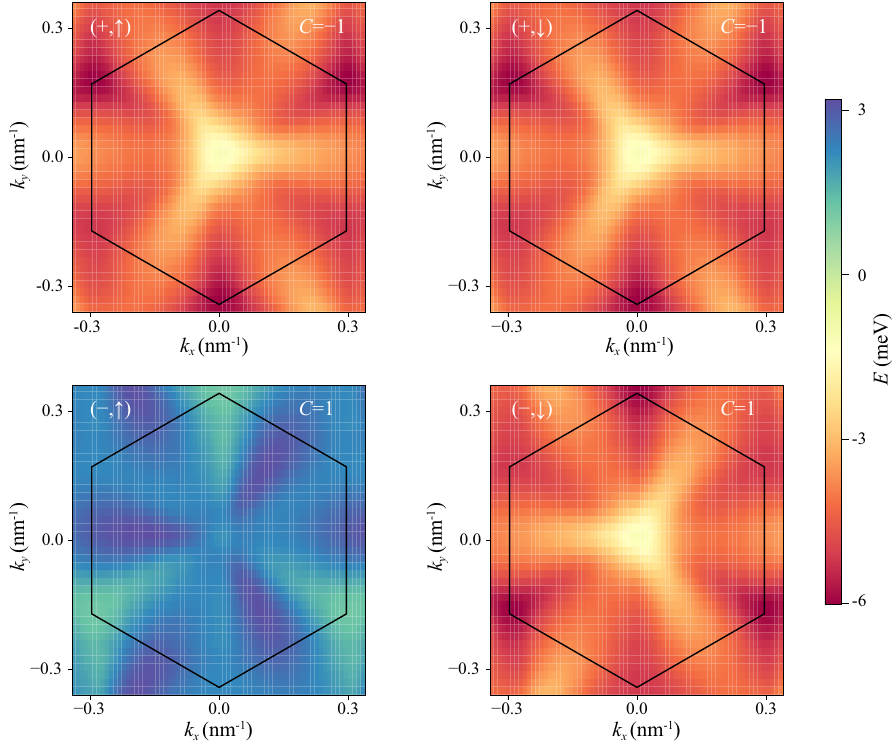}
  \end{center}
\caption{The four upper fat bands labeled by the valley-spin index $(\tau,s)$ are shown for the insulating valley-spin-polarized state obtained from the Hatree-Fock approximation. The Chern numbers of the flat bands are displayed. The black hexagon encloses the moir{\' e} BZ. 
} 
  \label{fig3}
\end{figure}

We then study the effect of current and magnetic field on the electron {dispersion}. The current is introduced through the Lagrange multiplier as before and the Hamiltonian becomes
\begin{equation}\label{H}
\mathcal{H} =\mathcal{H}_0 - \sum_{k,\tau,s}\left(\lambda_s \partial_{k_\alpha} \varepsilon_{k,\tau} + g\mu_B s_z B  + M_{\bm{k},\tau}B  \right) c_{k,\tau,s}^\dagger c_{k,\tau,s},
\end{equation}
where $\lambda_s$ denotes the Lagrange multiplier for the current with spin $s$ and along the $\alpha$ direction, $s_z=\pm1/2$ is the spin quantum number, and $B$ represents the perpendicular magnetic field. $M_{\bm{k},\tau}$ is the {$\bm{k}$-resolved} orbital {magnetization}
\begin{equation}
M_{\bm{k},\tau} = i\frac{e}{2\hbar}\Bra{\nabla_{\bm{k}}\psi_{\bm{k},\tau}} \times ( {2\mu_0} - \varepsilon_{\bm{k},\tau} - \mathcal{H}_{\bm{k},\tau}) \Ket{\nabla_{\bm{k}}\psi_{\bm{k},\tau}} \cdot \hat{\bm{e}}_z,
\end{equation}
{{that is} generated by the self rotation of Bloch states {and includes also the boundary contribution} \cite{PhysRevB.53.7010,PhysRevLett.95.137204,PhysRevB.74.024408,RevModPhys.82.1959}.} {Here $\mu_0$ is the chemical potential corresponding to the 3/4 filling of the bare dispersion, and $\hat{\bm{e}}_z$ is a unit vector perpendicular to TBG.}
The orbital magnetic moment is spin-degenerate and {valley-contrasting.} {Thus the symmetry breaking state at 3/4 filling has the polarized orbital magnetic moment, and exhibits orbital ferromegnetism.}
To be concrete, we focus on the current along the $x$ direction {and take the valley-spin-polarized ground state in Fig.\ref{fig3} as the initial state.  Because the flat bands with spin down are fully filled, the current can only be conducted by the spin up bands. Hence we fix $\lambda_\downarrow=0$ in the following. In this case, the Hatree-Fock approximation yields the self-consistent equation in the same form as Eq. (\ref{sf}) but with the replacement $\varepsilon_{\bm{k},\tau} \rightarrow \varepsilon_{\bm{k},\tau} - \lambda_s \partial_{k_x} \varepsilon_{\bm{k},\tau} - g\mu_B s_z B  - M_{\bm{k},\tau}B$.  
}

\begin{figure}[t]
  \begin{center}
  \includegraphics[width=8.5 cm]{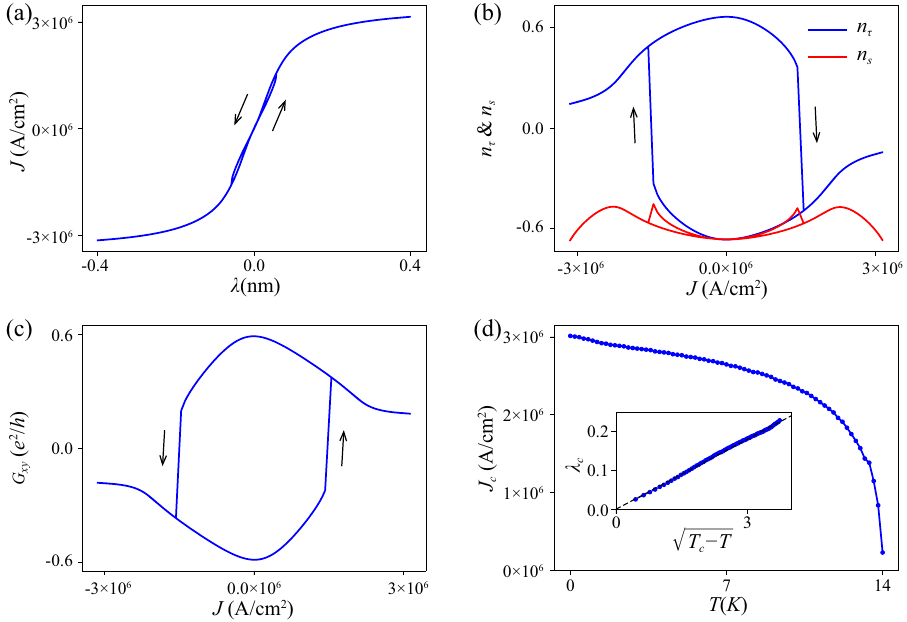}
  \end{center}
\caption{\ys{(a) Current density $J$ as a function of $\lambda$ at $T=13K$. (b) and (c) The corresponding order parameters  $n_{\tau,s}$ and Hall conductance $G_{xy}$ as a function of $J$. (d) The dependence of critical current density $J_c$ on $T$. The inset shows $\lambda_c$ vs $\sqrt{T_c-T}$.}  
} 
  \label{fig4}
\end{figure}

\begin{figure}[b]
  \begin{center}
  \includegraphics[width=6.5 cm]{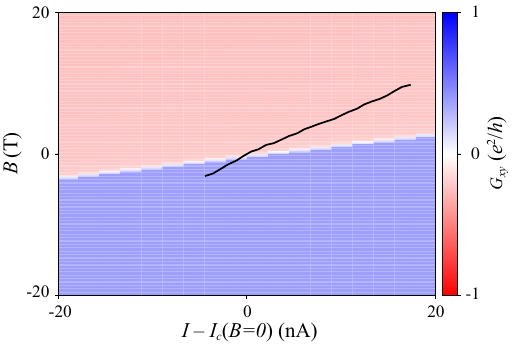}
  \end{center}
\caption{\ys{Density plot of the Hall conductance $G_{xy}$ as a function of current $I$ and magnetic field $B$. The positive and negative $G_{xy}$ are for opposite valley polarization. The black line is the phase boundary extracted from Ref. \cite{youngscience} that  separates different valley-polarized states. Here we offset $I$ by the critical current $I_c$ at $B=0$ to make a direct comparison of the phase boundaries.}}
  \label{fig5}
\end{figure}

We first focus on the effect of current on VP and Hall conductance in the absence of magnetic field $B=0$. The analysis in Eq. \eqref{eq2} reveals that the switching of VP by currents is of the first order phase transition, and therefore hysteresis is expected. We sweep {the} current by changing $\lambda$ continuously {and solve the corresponding self-consistent equations}. \ys{The current density through the TBG is $J = \frac{e}{Ah} \sum_{\bm{k},\tau,s} \frac{\partial \varepsilon_{\bm{k},\tau}}{\hbar\partial k_x} f(E_{\bm{k},\tau,s})$, where $h=0.6$ nm is the thickness of TBG. In Fig. \ref{fig4}(a), we show that $J$ increases monontonically with $\lambda$.}
The results of $n_\tau$, $n_s$, and $G_{xy}$ 
\ys{at $T=13K$} as a function of \ys{$J$} are displayed in Figs. \ref{fig4}(b) and \ref{fig4}(c){, where the} {hysteretic} behavior {of VP and Hall conductance appears as expected}. Start with the initial state in Fig. \ref{fig3} with $n_\tau\approx 0.6$ and $G_{xy}\approx -0.6 e^2/h$ (here $n_\tau$ and $G_{xy}$ are not saturated because of the nonzero $T$.), and then increase $J$, both $n_\tau$ and $G_{xy}$ change sign through a sharp jump at the critical current density $J_c$ (and the corresponding $\lambda_c$). If we remove the current {(by setting $\lambda=0$)} after the jump, the system then relaxes into an insulating state with opposite $n_\tau$ and $G_{xy}$. There is no switching of spin polarization because the bands with down spin remain fully occupied. \ys{Therefore, our results demonstrate that the VP and topology can be switched by a current larger than $J_c$.} \ys{The results for $T=0$ are presented in Ref. \cite{supplement}.}

{Here $\lambda_c$ and $J_c$ are determined by the energy barrier separating different valley-polarized states.}
As $T$ increases, the energy barrier 
decreases, and hence $\lambda_c$ and $J_c$ decreases with $T$. The dependence of $\lambda_c$ and $J_c$ on $T$ are shown in Fig. \ref{fig4}(d). {As} can be derived from Eq. \eqref{eq2}, we have the following scaling relations $\mathcal{F}\sim (T_c-T)^2$, $\delta\sim \sqrt{T_c-T}$, and $\alpha_3\sim \lambda$. Therefore $\lambda_c\sim\sqrt{T_c-T}$ consistent with the numerical results {shown in the inset of Fig. \ref{fig4}(d)}.

The energy barrier is affected by the magnetic field through the Zeeman coupling with spin and orbital magnetic moments in Eq. (\ref{H}). In TBG, the orbital magnetic moment is dominant and is valley-contrasting. The energy gap of the insulating valley-spin-polarized state scales linearly with $B$. {To study the dependence of critical current $I_c$ on $B$, we show} $G_{xy}$ as a function of current $I$ for different $B$ at \ys{$T=13K$}  in Fig. \ref{fig5}. The switching of VP and topology at $I_c$ is featured by the jump of $G_{xy}$ where $G_{xy}$ changes sign. \ys{$I_c$ as a function of $B$} indicated by the black line in Fig. \ref{fig5} \ys{is extracted from Ref. \cite{youngscience}. Both the experimental result and our numerical simulation show a linear dependence of $I_c$ on $B$.}

{\emph{Discussion and summary.}---}\ys{Here we compare our theory to the other proposals \cite{youngscience, he_giant_2019,huang_current_2020}. In Ref. \cite{youngscience}, the authors attributed the switching of VP due to the chiral edge current, where the asymmetric edges tilt the free energy difference for two otherwise degenerate VP states. However, the switching of the VP requires overcoming the bulk energy barrier. For a large system size with a single domain, the edge contribution is not enough to overcome the bulk energy barrier \cite{supplement}. One may argue that the switching involves multiple domains. In this picture, the domain walls must percolate the whole system in order to conduct current. In Ref. \cite{he_giant_2019}, He \emph{et al.} argued that strain due to the substrate can break the $C_3$ symmetry down to $C_1$ symmetry. The magnetoelectric effect then can allow to change the magnetization by electric field, and switch the VP. This requires an electric field throughout the system and finite quasiparticle lifetime, and therefore does not apply directly to the QAHE states. In Ref. \cite{huang_current_2020}, the authors considered domain walls between different VP. They argued that the electric current can couple to the domain wall described by the valley pseudospin and drive the domain wall into motion. The motion of domain wall results in the expansion of domains with one VP at the expense of the domains with the opposite VP, and eventually flips the VP.

In our approach, we highlight the effect of the bulk current. A bulk current in the QAHE state is possible either due to thermally activated quasiparticle or when the electric field exceeds a threshold value determined by the gap \cite{supplement}. Our scenario can be tested experimentally by studying devices with a single domain. The single domain state can be prepared by polarizing the devices with external magnetic field at low temperature \cite{youngscience}. One then measures the longitudinal current-voltage curve and the corresponding Hall conductance, and compares to the theoretical prediction in Fig. \ref{fig4}. 

Our theory can be generalized to the multiple domains case. The mechanism proposed here offers a way to change the energy for different VP domains by current, which facilitates the nucleation of the energetically favored VP in the presence of fluctuations. Therefore, the threshold switching current identified for a single domain when the energy barrier vanishes completely corresponds to the superheating/supercooling field.
}

In summary, we propose a mechanism of the switching of {VP} and topology in TBG by bulk electric currents. The current causes the redistribution of electron occupation, that lifts the degeneracy and even overcomes the energy barrier between different valley-spin-polarized states through the Coulomb interaction. Our theory can be generalized to other strongly correlated two dimensional materials with valley polarization.

\begin{acknowledgements}

{\emph{Acknowledgements.}}---The authors thank Fengcheng Wu and {Di Xiao} for helpful discussion. Computer resources for numerical calculations were supported by the Institutional Computing Program at LANL. This work was carried out under the auspices of the U.S. DOE NNSA under contract No. 89233218CNA000001 through the LDRD Program, and was supported by the Center for Nonlinear Studies at LANL and Institute for Materials Science (IMS) at LANL through IMS Rapid Response. S. Z. L. was also supported by the U.S. Department of Energy, Office of Science, Basic Energy Sciences, Materials Sciences and Engineering Division, Condensed Matter Theory Program.

\end{acknowledgements}

\bibliography{references}

\widetext
\clearpage
\begin{center}
\textbf{\large Supplemental Material: Switching of valley polarization and topology in twisted bilayer graphene by electric currents}
\end{center}
\setcounter{equation}{0}
\setcounter{figure}{0}
\setcounter{table}{0}
\setcounter{page}{1}
\makeatletter
\renewcommand{\theequation}{S\arabic{equation}}
\renewcommand{\thefigure}{S\arabic{figure}}
\renewcommand{\bibnumfmt}[1]{[S#1]}
\renewcommand{\citenumfont}[1]{S#1}

\twocolumngrid

\ys{
\subsection{I. Lagrangian multiplier for current}
The simplest way to derive the Lagrangian multiplier term in Eq. (1) in the main text is considering a torus geometry with a threading magnetic flux. The torus geometry preserves the translational invariance in both directions and the momentum is a good quantum number. The magnetic flux generates a current. The flux can be introduced into the Hamiltonian through the standard Peierls substitution $\bm{k}\rightarrow \bm{k}{+}e \bm{A}/\hbar $. The single particle Hamiltonian becomes
\begin{align}
    \mathcal{H}_0=\varepsilon(\bm{k}{+}e \bm{A}/\hbar ) c_k^\dagger c_k\approx \left[ \varepsilon(\bm{k}){+}\frac{e \bm{A}}{\hbar } \frac{\partial \varepsilon_{\bm{k}}} {\partial \bm{k}}\right]  c_k^\dagger c_k, 
\end{align}
which has the same form as the Lagrangian multiplier term in the main text. The corresponding current is given by $\bm{J}=-\partial\mathcal{H}/\partial \bm{A}$. In the absence of dissipation, a stationary current can be induced by a static flux. When dissipation is present, a time dependent flux is required to generated a stationary current. The current dissipates over a time scale associated with the quasiparticle lifetime $t_0$. Heuristically, we can take $\bm{A}=\bm{E_e} t_0$, where $\bm{E}_e$ is the electric field. {Therefore, the Lagrangian multiplier in Eq. (1) is $\lambda=-e t_0 E_e/\hbar$.}

The meaning of the Lagrangian multiplier can also be understood intuitively using the the Boltzmann transport equation 
\begin{equation}
\frac{\partial f}{\partial t} + \frac{\partial f}{\partial \bm{r}} \cdot \bm{v} + \frac{\partial f}{\hbar\partial \bm{k}}\cdot (-e)\bm{E}_e = - \frac{f-f_0}{t_0},
\end{equation}
under the relaxation time approximation \cite{solid}. Here $f_0$ is the Fermi-Dirac function and $f$ is the distribution function in the presence of the bias voltage. For an uniform electric field $\bm{E}_e$, the distribution function in the stationary state becomes 
\begin{equation}
f(\varepsilon_{\bm{k}}) = f_0(\varepsilon_{\bm{k}}) + \frac{et_0}{\hbar}\frac{\partial f_0(\varepsilon_{\bm{k}})}{\partial \varepsilon_{\bm{k}}}  \frac{\partial \varepsilon_{\bm{k}}}{\partial \bm{k}} \cdot \bm{E}_e \approx f_0\left( \varepsilon_{\bm{k}} + \frac{e t_0}{\hbar} \frac{\partial \varepsilon_{\bm{k}}}{\partial \bm{k}} \cdot \bm{E}_e \right).
\end{equation}
Then we can identify the Lagrangian multiplier as  $\lambda = {-e t_0 E_e}/{\hbar}$, {that is consistent with the derivation above.}
}

\subsection{II. Explicit calculation of the toy model}
Here we provide explicit calculation of the toy model using the dispersion relation $\varepsilon_{k,-}=\cos(2k+\pi)-\mu$ for $-\pi\le k \le -\pi/2$ and $\varepsilon_{k,-}=\cos[\frac{2}{3} (k+\pi/2)]-\mu$ for  $-\pi/2\le k \le \pi$, and $\varepsilon_{k,+}=\varepsilon_{-k,-}$. 
The results for $\Delta_\pm$ and current $J$ as a function of $\lambda$ are displayed in Fig. \ref{figx1}. The switching of VP occurs only for a positive $\lambda$ because we start with the initial state shown in Fig. 1(a). For  $V=3$ and $\mu=1.5$, the switching of VP (featured by the crossing of $\Delta_\pm$) happens simultaneously with the insulator-to-metal transition, as shown in Figs. \ref{figx1}(a) and \ref{figx1}(b). When the gap is increased by choosing $V=5$, the system first becomes a metal, and then switches the VP upon further increasing $\lambda$, see Figs. \ref{figx1}(c) and \ref{figx1}(d).

\begin{figure}[t]
  \begin{center}
  \includegraphics[width=\columnwidth]{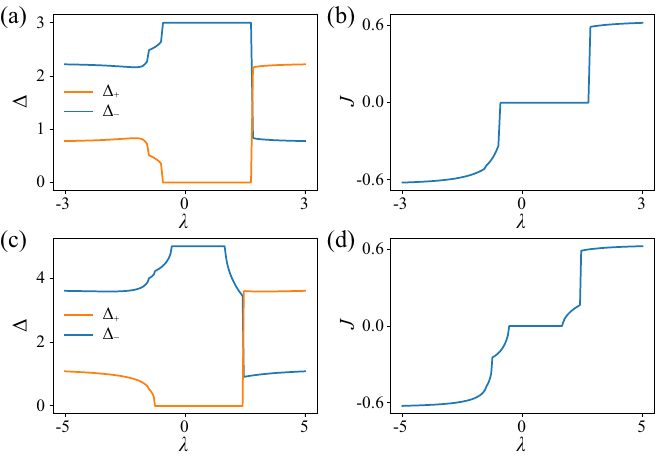}
  \end{center}
\caption{ (a) and (c) $\Delta_\pm$  as a function of $\lambda$ for $V=3$ and $5$. The corresponding current $J$ are shown in (b) and (d), respectively.}
  \label{figx1}
\end{figure}

The free energy of the system is 
\begin{align}
\mathcal{F}=-{k_BT}\sum_{k,\tau}\ln\left[1+\exp(-E_{k,\tau}/k_BT)   \right]-\frac{\Delta_+\Delta_{-} }{V}.
\end{align}
When the VP order parameter $\delta=(\Delta_+-\Delta_-)/2$ is small, we can expand $\mathcal{F}$ in terms of $\delta$ as shown in Eq. (2). The coefficients are
\begin{align}
\mathcal{F}_0=-{k_BT}\sum_{k,\tau}\ln\left[1+\exp[-(\varepsilon_{k,\tau}-\lambda\partial_k\varepsilon_{k,\tau}+\rho)/k_BT]   \right]-\frac{\rho^2}{V},
\end{align}
\begin{align}
\alpha_1=-\sum_{k,\tau}\frac{\tau}{ 1+\exp(g) },
\end{align}
\begin{align}
\alpha_2=-\frac{1}{2}\sum_{k,\tau}\frac{1}{ 2k_BT+2k_BT\cosh(g) }+\frac{1}{V},
\end{align}
\begin{align}
\alpha_3=-\frac{1}{6}\sum_{k,\tau}\frac{2\tau \mathrm{csch}^3(g)\sinh^4(g/2)}{ k_B^2T^2 },
\end{align}
\begin{align}
\alpha_4=-\frac{1}{24}\sum_{k,\tau}\frac{[-2+ \cosh(g)]\mathrm{sech}^4(g/2)}{ 8k_B^3T^3 },
\end{align}
with $\rho=(\Delta_++\Delta_-)/2$ and $g=(\varepsilon_{k,\tau}-\lambda\partial_k\varepsilon_{k,\tau}+\rho)/k_BT$. In the absence of current $\lambda=0$, $\alpha_1=\alpha_3=0$ as required by the $Z_2$ symmetry associated with the VP. One can show explicitly that $\alpha_{1,3}(\lambda)=-\alpha_{1,3}(-\lambda)$ and $\alpha_{2,4}(\lambda)=\alpha_{2,4}(-\lambda)$.

The free energy as a function of $\delta$ for different $\lambda$ and with $V=3$ and $\mu=1.5$ is shown in Fig. \ref{figx2}. The presence of a nonzero $\lambda$ lifts the degeneracy between the two valley-polarized states. As a consequence, one valley-polarized state becomes metastable, while the other becomes the global ground state. Upon further increasing or decreasing $\lambda$, the metastable state becomes unstable and the switching of VP happens.

\begin{figure}[t]
  \begin{center}
  \includegraphics[width=6 cm]{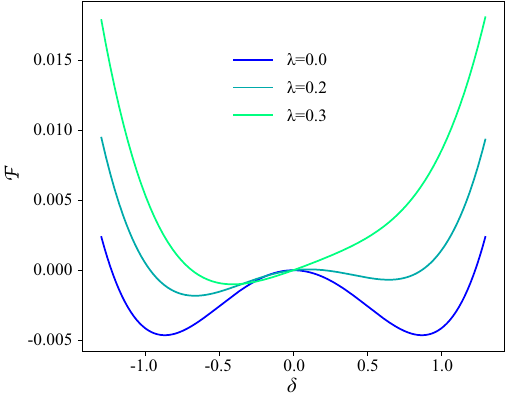}
  \end{center}
\caption{Free energy as a function of $\delta$ for different $\lambda$ and $V=3$. }
  \label{figx2}
\end{figure}



\ys{
\subsection{III. Dependence on the gate distance}

The gate distance $d$ controls the screening of the Coulomb interaction. As $d$ decreases, the screened Coulomb potential becomes weaker. {When the Coulomb interaction is not over-screened}, the valley-spin-polarized state in Fig. \ref{fig3} remains the ground state and the QAHE persists. The energy gap of the valley-spin-polarized QAH state increases with $d$, as shown in Fig. \ref{figx3}. Since the robustness of the QAH state depends on the gap, the critical temperature for the QAHE and the critical current for the switching of valley polarization both increase with $d$.

\begin{figure}[htb]
  \begin{center}
  \includegraphics[width=5 cm]{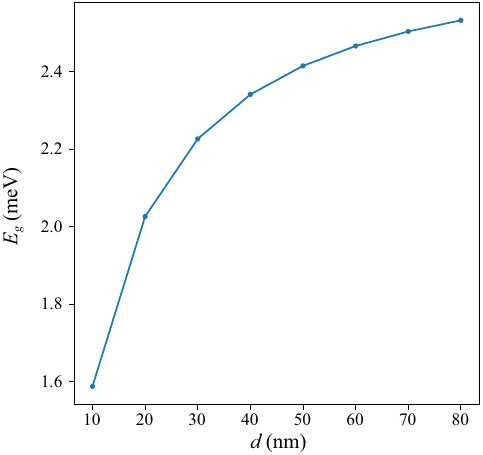}
  \end{center}
\caption{Energy gap of the valley-spin-polarized QAH state as a function of the gate distance $d$. }
  \label{figx3}
\end{figure}

\subsection{IV. Switching of valley polarization at zero temperature}

In Fig. \ref{figx6}, we show the current density $J$, order parameters $n_\tau$ and $n_s$, and Hall conductance $G_{xy}$ as a function of $\lambda$ at $T=0$ K. The results exhibit that a critical voltage is required to induce a current before the switching of the Hall conductance. Here the voltage $V_e=E_eL\propto \lambda$, where $L$ is the length of the system. The critical voltage is determined by the insulating gap. Above the critical voltage, electrons gain enough energy to overcome the insulating gap to conduct current.}

\begin{figure}[t]
  \begin{center}
  \includegraphics[width=8 cm]{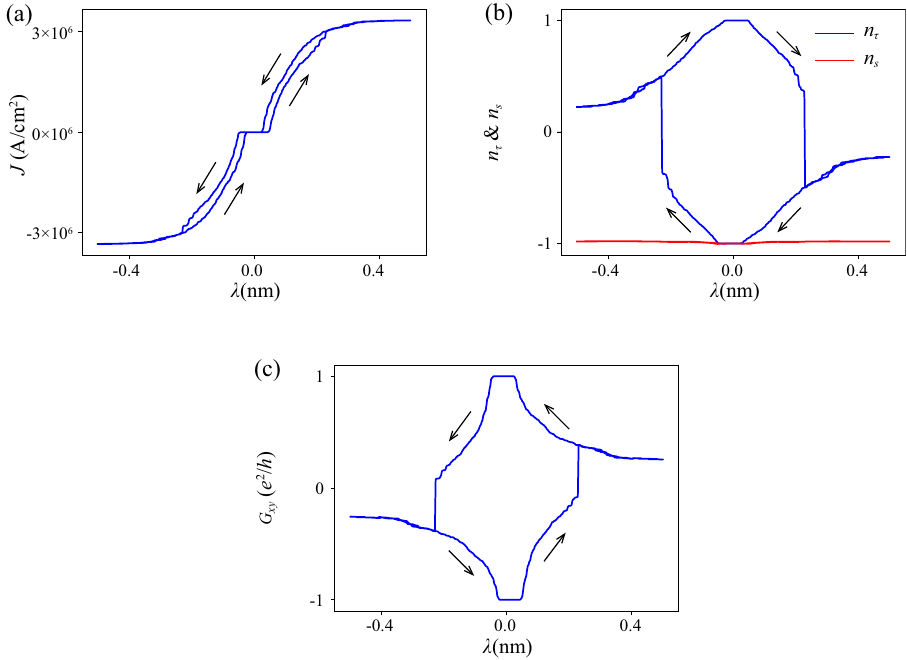}
  \end{center}
\caption{(a)-(c) Current density $J$, order parameters $n_\tau$ and $n_s$, and Hall conductance $G_{xy}$ as a function of $\lambda$ at $T=0K$. The arrows indicate the direction of  evolution of the hysteresis loops.}
  \label{figx6}
\end{figure}

\subsection{V. Estimate the edge current contribution in overcoming the energy barrier}

Here we estimate the edge current contribution in overcoming the energy barrier between different ground states with opposite VP. Because of the large unit cell size at a small twist angle, it is hard to access the edge states of the TBG. Instead, we use the edge states of the Haldane model to mimic that of TBG.   Explicitly, we consider the Haldane model  described by the Hamiltonian \citep{haldane} 
\begin{equation}
\begin{split}
\mathcal{H}_H&=2t_2\cos\phi \sum_{i=1}^3 \cos(\bm{k}\cdot\bm{b}_i) \sigma_0 \\
&+ t_1\sum_{i=1}^3\left[ \cos(\bm{k}\cdot\bm{a}_i)\sigma_x + \sin(\bm{k}\cdot\bm{a}_i)\sigma_y  \right] \\
&+ \left[ M-2t_2\sin\phi\sum_{i=1}^3 \sin(\bm{k}\cdot\bm{b}_i)  \right] \sigma_z,
\end{split}
\end{equation}
on a honeycomb lattice whose lattice constant is set to unity. Here $t_n$ is the hopping energy between the $n$th nearest neighboring (NN) lattice sites, $\sigma_0$ is the $2\times 2$ identity matrix, and $\sigma_{x,y,z}$ are the Pauli matrices acting on the sublattice degree of freedom. $\bm{a}_{1,2,3}$ are the three vectors connecting NN lattice sites, and $\bm{b}_{1,2,3}$ are primitive lattice vectors related by 3-fold rotation. $M$ is the sublattice potential breaking the inversion symmetry, and $\phi$ is the phase factor associated with the second NN hopping. The topological phase of the model is determined by $M$ and $\phi$, as shown by the phase diagram in Fig. \ref{Haldane}(a). By choosing $t_2=0.07t_1$ and $M=3t_2$, we show the free energy as a function of $\phi$ of the Haldane model at half filling and zero temperature in Fig. \ref{Haldane}(b). Apparently, there are two degenerate energy minimum with opposite Chern number $C=\pm1$ at $\phi=\pm\pi/2$, that is similar to the TBG. Therefore, by taking $\phi$ as an internal order parameter (playing the same role of $\delta$), we can use the Haldane model to mimic the free energy profile and topology of the TBG.

\begin{figure}[t]
  \begin{center}
  \includegraphics[width=8.5 cm]{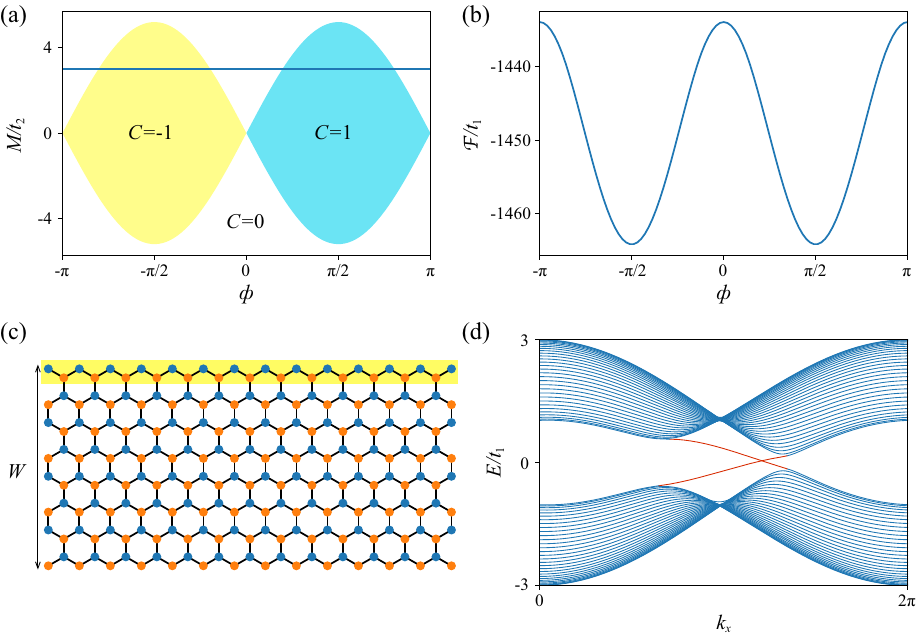}
  \end{center}
\caption{(a) Phase diagram of the Haldane model. The Chern numbers of different phases are shown. (b) Free energy of the Haldane model along the straight line in (a) at half filling and zero temperature. (c) A strip of honeycomb lattice with width $W$ and zigzag edges. (d) Energy spectrum of the Haldane model on the honeycomb lattice strip. } 
  \label{Haldane}
\end{figure}

To study the edge current contribution, we consider a stripe of honeycomb lattice with zigzag edges along the $x$ direction, as shown in Fig. \ref{Haldane}(c). The edge asymmetry can be introduced by adding an onsite potential $V$  to the upper edge, which is highlighted in Fig. \ref{Haldane}(c). For $V=2t_2$, the energy spectrum of the strip  with width $W=25\sqrt{3}$ (that contains 50 unit cells in the cross section) is shown in  Fig. \ref{Haldane}(d). Here the edge states traversing the the bulk energy gap are marked by red color. In the presence of a current along the strip, the system is described by $\mathcal{H}_H- \sum_n \lambda \partial_{k_x} \varepsilon_{k,n}$
where $\varepsilon_{k,n}$ represents the bare dispersion of the strip [see Fig. \ref{Haldane}(d)] and $n$ is the band index. The current lifts the degeneracy of the two energy minima at $\phi=\pm\pi/2$, as shown in Fig. \ref{spectrum}(a). The energy barrier between the two minima vanishes at a critical $\lambda$, i.e. $\lambda_c\approx 1.25$ in this case, that corresponds to the switching of VP and topology in TBG. To show the edge current contribution in overcoming the energy barrier, we study the electron occupation in the conductance and valence bands at $\lambda=\lambda_c$. As a reference, we show the electron occupation in the absence of current for $\lambda=0$ in Fig. \ref{spectrum}(b), where the conductance (valence) bands are fully empty (occupied) at zero temperature. The two branches of counter-propagating edge states are partially filled. In the presence of the critical current for  $\lambda=\lambda_c$, the electron occupation is shown in Fig. \ref{spectrum}(c). Apparently, the branch of edge states propagating along the current direction is fully filled, while the other branch is empty. Moreover, there are electrons pumped from the valence bands to conductance bands as in the TBG. Namely, the edge current alone is not enough to overcome the energy barrier in this case. The pumped electrons occupy the the bottom of conduction bands where the density of states is high. Therefore, the bulk current plays a dominant role in overcoming the energy barrier for a large system. The edge current can be important if the sample size is small since the energy barrier grows linearly in the sample volume. However, in experiments [6,7], the device is in micron size, that contains hundreds of moir{\' e} unit cells (considering the lattice constant of the small-angle TBG is of the order of $10$ nm) in the cross section. This fact motivates us to study the switching of VP and topology in TBG by electric currents in the bulk.

\begin{figure}[hbt]
  \begin{center}
  \includegraphics[width=8.5 cm]{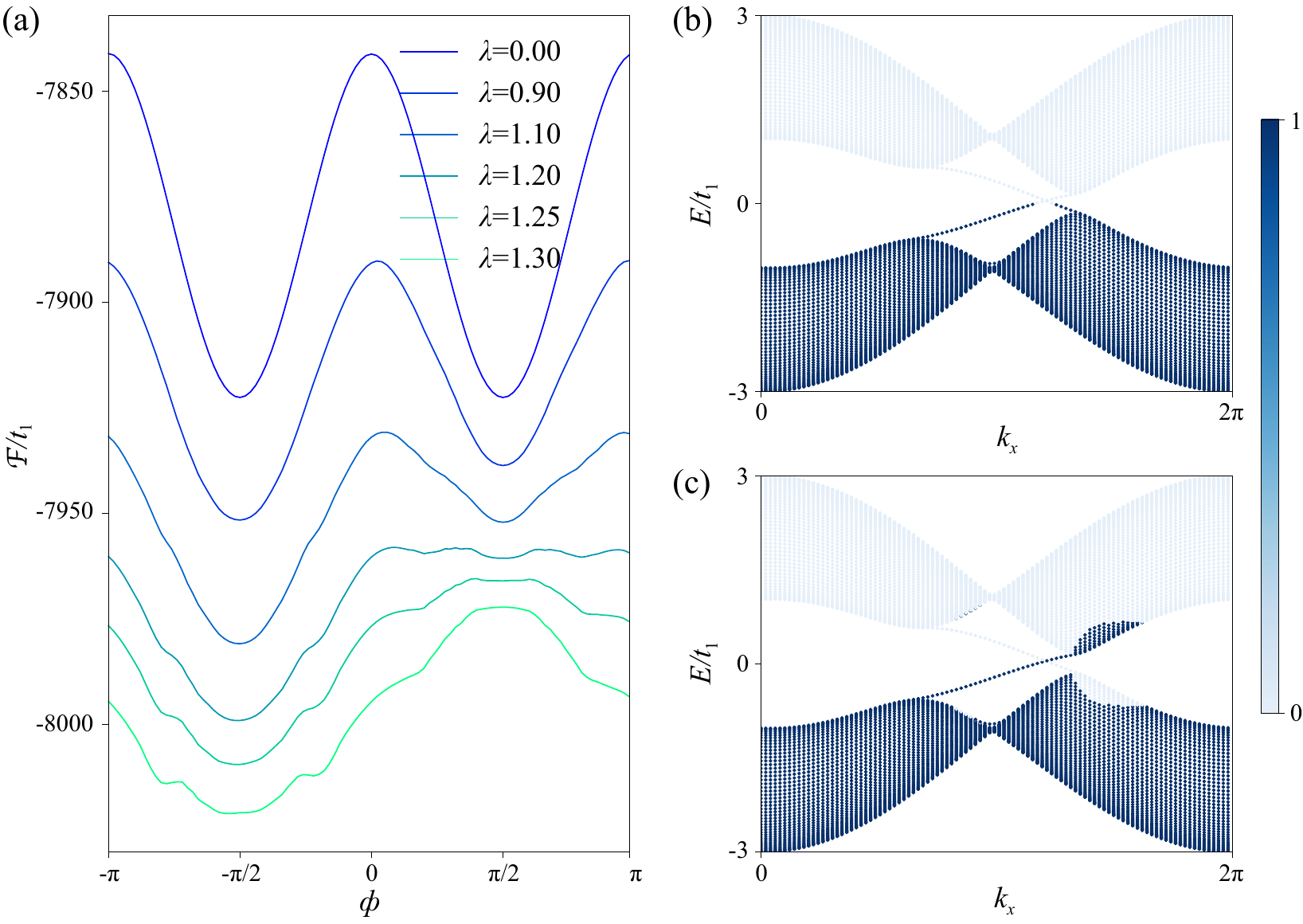}
  \end{center}
\caption{(a) Free energy of the Haldane model  as a function of $\phi$ for different $\lambda$. (b) and (c) Electron occupation for $\lambda=0$ and for $\lambda=1.25$, respectively.} 
  \label{spectrum}
\end{figure}

\end{document}